\documentclass{ws-procs975x65}
\usepackage{graphicx}
\def\beq{\begin{equation}}
\def\eeq{\end{equation}}

\begin{document}

\title{The transparency of the universe for very high energy gamma-rays}
\author{Dieter Horns}

\address{Department of Physics, University of Hamburg,\\
D-22761 Hamburg,Germany\\
E-mail: dieter.horns@desy.de}

\begin{abstract}
 The propagation of very high energy gamma-rays ($E>100$~GeV) over cosmological
distances is suppressed by pair-production processes with the ubiquitous
extra-galactic soft photon background, mainly in the optical to near infra-red. The
detailed spectroscopy of gamma-ray emitting blazars has revealed the signature
of this absorption process leading to a meaningful measurement of the
background photon field which is linked to the star-forming history of the
universe. Deviations from the expected absorption have been claimed in the past.
Here the status of the observations is summarized, an update 
on the search for the persisting anomalous transparency is given and discussed.
\end{abstract}


\bodymatter


\section{Introduction and Context}
\begin{figure}
	\begin{minipage}[t]{0.48\linewidth}
		\includegraphics[width=\linewidth]{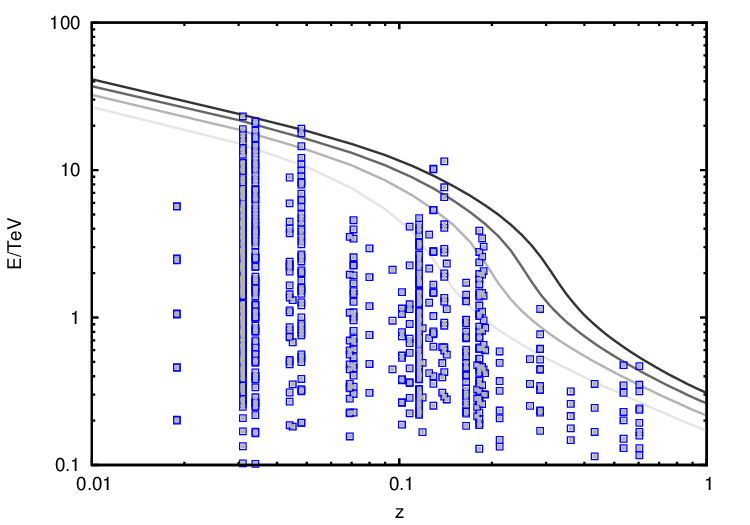}
		\caption{Isocontours of optical depth (from bottom to top:
			$\tau=1,2,3,4$) 
			in the plane of 
			red shift and observed photon energy (absorption in the extra-galactic
			background light from a backward evolutionary model\cite{franceschini08}).
			The individual points are from a collection of spectral measurements
			from gamma-ray sources similar to the one collected in 
	        \cite{biteauwilliams}. \label{iso} }
	\end{minipage}
	\hspace{0.02\linewidth}
       \begin{minipage}[t]{0.48\linewidth}
   \includegraphics[width=\linewidth]{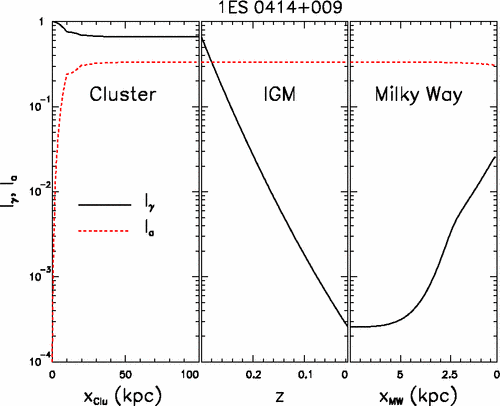}
   \caption{Photon ($I_\gamma$) and ALP ($I_a$) number evolution while propagating through 
\label{Bild1}
the magnetized medium of the galaxy cluster environment, the inter-galactic
medium (including photon absorption processes, but no mixing considered) 
and finally through the
magnetized halo of the Milky Way. Reprinted figure
 with permission from \cite{hornsmirizzi} Copyright (2012) by the 
American Physical Society.}
 \end{minipage}
\end{figure}

The  current generation of ground based imaging air Cherenkov telescopes 
has been sufficiently sensitive to discover astrophysical 
gamma-ray emission in the energy range above
100 GeV from currently $>60$ active galactic nuclei (at the
time of writing). The current record-holder
as the most distant known gamma-ray source is observed at a red shift 
of $z=0.944$.\cite{ATEL6349}  At this distance, the effect of pair production
becomes important at energies above 100~GeV leading to the energy-dependent 
absorption of the primary beam (see Fig.~\ref{iso}).  
For objects at a red shift of $0.01$, absorption is only of relevance at
photon energies above 10 TeV. \\
This energy-dependent  effect is well known.\cite{Nikishev} However, the uncertainties (by roughly 
a factor of two) on
the amount of background light translate directly to  uncertainties 
on the predicted optical depth $\tau$, which widely changes the resulting attenuation $\exp(-\tau)$.  \\
Additional effects related to the production of secondary photons
in inter-galactic cascades may have an impact on the observed energy spectrum. \cite{gould1978,essey}
The observable (secondary) flux depends 
on the intervening magnetic field and plasma properties which 
can lead to quenching of the cascade. \\
Finally, phenomena beyond the standard model of particle physics 
could modify this picture in characteristic ways (see below). \\
Before turning
to the modifications of the pair production processes, I summarize the current
state of knowledge on the extra-galactic background light (EBL) drawn from
the imprint of absorption in the observed gamma-ray spectra (for an extended review, see also [\refcite{horns_agnieszka}]).
\section{Measurement of the Extra Galactic Background Light (EBL)}
The direct observation of the extra galactic background light (EBL) is strongly affected
by the dominating 
\textit{foreground} emission from re-processed light in the interplanetary and interstellar medium
(see e.g., [\refcite{Leinert}]).
 The indirect approach to infer the amount of EBL present in the universe
 through the energy dependent imprint of absorption in gamma-ray spectra does not suffer
 from this uncertainty. Furthermore, it is in principle possible
 to follow the evolution of the EBL by measuring sources at various red shifts.\\
While in the past, the observed spectra were not determined with sufficient accuracy to 
derive the amount of absorption, current measurements of various bright sources
(predominantly during flaring episodes, \cite{hess2155}),  
as well as the combination of available source spectra have been used to
estimate the absorption and subsequently provide a measure of the 
amount of EBL present.\cite{biteauwilliams}\\
Intrinsic to these analyses is however the ignorance of the intrinsic source
spectra which is leading to systematic uncertainties on the reconstructed EBL.
The assumption used, e.g., that the source spectrum is a power-law,  may be
over-simplifying: specifically for source spectra observed during flaring episodes,
transient spectral components may be present at narrow energy intervals.\cite{lefa2011} 
Combining the observations with lower-energy measurements
with the pair-production telescope Fermi-LAT provides additional information on the un-absorbed energy spectrum and the presence of a spectral
cut-off.  \\
An interesting feature of the intrinsic spectra 
is an apparent dependence with red shift: the intrinsic source
spectra show a hardening with increasing red shift which is not obviously 
linked with source
characteristics. This trend is also 
present in a recent analysis,\cite{biteauwilliams} and has been 
pointed out already in previous works.\cite{roncadelli, grisha}
The apparent hardening of the gamma-ray spectra with increasing red shift
could be explained by mixing photons with a light pseudoscalar boson.\cite{roncadelli}
Additional studies indicate that the excess emission is mostly present in the optically thick
part of the spectrum.\cite{horns}
\section{Indications for anomalous transparency: An update}
\begin{figure}
	\begin{minipage}[t]{0.48\linewidth}
		\includegraphics[width=\linewidth]{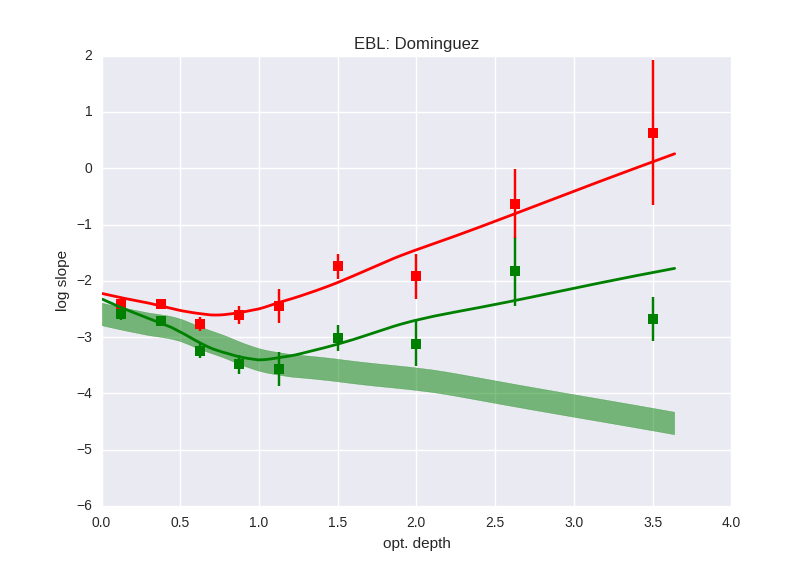}
		\caption{The logarithmic slope $\alpha$ (lower green points) 
			and $\alpha^\prime$ (upper red points) in bins of optical
		depth using the EBL model\cite{Dom}\label{fig:dom}.
		The solid lines are non-parametric estimators,\cite{Cleveland92}
		the band indicates the expected slope (assuming an intrinsic index
	of $\alpha^\prime=-2.4$ as measured for small values of optical depth).}
	\end{minipage}
	\hfill
	\begin{minipage}[t]{0.48\linewidth}
	\includegraphics[width=\linewidth]{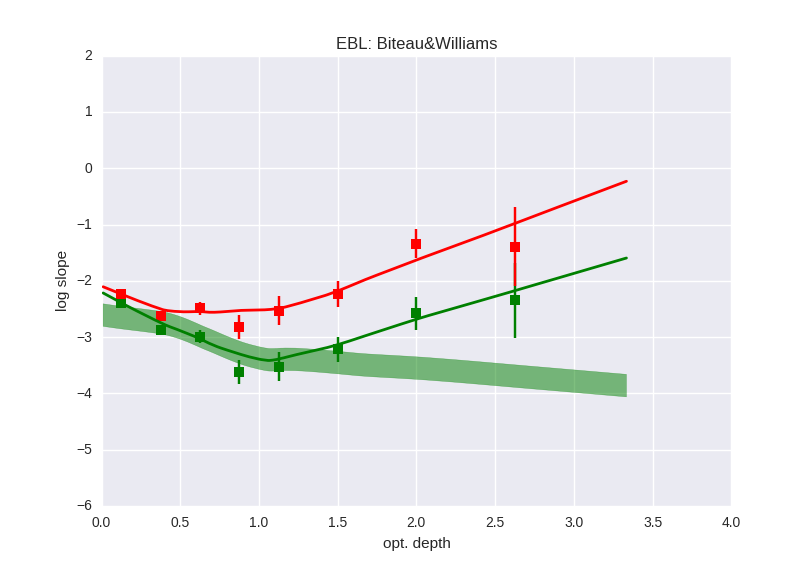}
	\caption{The average slopes $\alpha$ (observed) and $\alpha^\prime$
		(corrected for absorption) for 
		the EBL absorption model reconstructed from gamma-ray spectra.\cite{biteauwilliams}
	\label{fig:bw}}
	\end{minipage}
\end{figure}

The claim of an indication for anomalous transparency (pair production anomaly\cite{horns})
has recently been shown to (mostly) disappear,\cite{biteauwilliams} when 
increasing the data set and using an EBL derived from gamma-ray spectroscopy itself. The 
initial search for an
anomalous transparency was based upon a flux ratio which is sensitive to excess emission
when comparing the extrapolated energy spectrum with the observed value.\cite{horns} \\
Here, we introduce a new approach to characterise the spectral shape for different values
of optical depth.\\
The analysis is based upon the so-called \textit{gamma-ray cosmology} sample 
of blazars with known red shift.\cite{biteauwilliams} Instead of using 
the power-law index determined from the spectrum observed at small optical depth 
to compare with the measured values at large optical depth,\cite{horns,biteauwilliams}
a local measurement of the \textit{observed} logarithmic slope $\alpha_i$ 
between two differential flux measurements
$F_i(E_i) $ and $F_{i+1}(E_{i+1})$ ($E_{i+1}>E_i$) 
is used:
\begin{equation}
	\alpha_i = \frac{\ln(F_{i+1}) - \ln(F_i)}{\ln(E_{i+1})-\ln(E_i)},
\end{equation}
such that the \textit{intrinsic} logarithmic slope $\alpha_i^\prime$ is given by
\begin{equation}
	\alpha_i^\prime = \alpha_i + \Delta_i(z), 
\end{equation}
with 
\begin{equation}
	\Delta_i(z) = \frac{\tau_{i+1}(z)-\tau_i(z)}{\ln(E_{i+1})-\ln(E_i)},
\end{equation}
where $\tau_{i}=\tau(E_{i},z)$ is calculated according to 
an EBL model of choice. \\
Flux points with low statistical significance (and potentially large flux bias),ie., 
which differ by less than 1.5 times the flux uncertainty from a zero value are
excluded from the sample. An \textit{ad hoc} flux bias correction has been applied
to the remaining flux points, where
the flux $F$ with uncertainty $\sigma_F$ is corrected in the following way:
\begin{equation}
	F_\mathrm{cor} = F \left[ 1+ \left(\frac{F}{\sigma_F}\right)^\zeta \right]^{-1},
\end{equation}
with $\zeta\approx -2.5$ derived from a simple toy Monte Carlo assuming purely Gaussian
statistics (a similar procedure is outlined in [\refcite{2mass}]). \\
The resulting values of $\alpha$ and $\alpha^\prime$ are averaged in intervals of
$\tau$. In Figs.~\ref{fig:dom} and \ref{fig:bw}, the data are shown for two 
different types of absorption models. The \textit{observed} slope $\alpha$ 
(lower set of points in green) displays a power-law index of approximately $-2.4$ 
at small optical depth (in the first bin). Up to an optical depth of $\tau \approx 1$,
the \textit{observed} slope softens to an average value of $\alpha \approx -3.5$, consistent
with \textit{intrinsic} $\alpha^\prime\approx -2.4$. However, with further
increasing optical depth, the \textit{observed} slope $\alpha$ increases again until for 
$\tau\gtrsim 2$, $\alpha\approx -2.4$. \\
The result is very similar for both models and repeats also for other tested models of
gamma-ray absorption. 

\section{Discussion}
 The observations discussed above as well as the update on searches for anomalous transparency
 presented here indicate a puzzling feature in the observed gamma-ray spectra: at optical 
 depth $\tau\gtrsim 2$ the spectra have a similar spectral slope as at small optical depth,
 consistent with previous analysis which have considered the global (ie., over
 the entire observable energy range covered for each source) power-law 
 index.\cite{roncadelli,grisha}
\subsection{Modification of Gamma-Ray Propagation in the Presence of Light Pseudoscalars: ALPs}
 In the presence of transversal magnetic fields, gamma-rays can mix with 
 a light (mass in the range of $\mu$eV-neV) pseudoscalar fundamental bosons 
 leading to an attenuation of the 
energy spectrum above a critical energy which depends on the mass of these
particles (so-called \textit{ALPS}- axion-like particles; for a review see [\refcite{alps}]) 
as well as on the magnetic field.\cite{raffelt87}  \\
While the photonic part of 
the propagating beam is attenuated in the EBL,
the pseudoscalar fraction can in principle propagate largely unimpeded until reconverting, e.g.,
in the Galactic magnetic field (see Fig.~\ref{Bild1}, note, the mixing in the inter-galactic space was assumed
to be negligible). 
The mixing can at most convert $1/3$ of an unpolarized photon beam into an
ALPs beam, of which at most $2/3$ can reconvert into photon fields. Overall, 
in the most favorable conditions, a fraction of $1/3\cdot 2/3=2/9$ of the initial photon beam 
can re-appear through the effect of conversion and re-conversion. 
This component will be however dominating over the attenuated  beam
if the optical depth $\tau\gtrsim \ln(2)\approx 0.7$ (assuming no mixing in inter-galactic space,
the photon beam suffers absorption and is depleted to $2/3~\exp(-\tau)$; when
entering the re-conversion region in the halo magnetic field of the Milky Way, 
in maximum mixing, 1/3 of the photons will convert to ALPs, leaving $4/9~\exp(-\tau)+2/9$, where 
the last term is the re-converted ALPs part of the beam; in total $4/9 (\exp(-\tau)+1/2)$ 
can be observed and once $\tau>2$, the reconverted part of the beam will dominate). This is in qualitative agreement
with the observation of a hardening of the observed spectrum for the part of the spectra
observed at an optical depth $\gtrsim 1$ as discussed above. Note, a more
quantitative study has been carried out indicating that the required mixing to explain
the observations has to be close to maximum, still consistent with other constraints on
possible photon-ALPs coupling.\cite{horns}
\subsection{Modification of gamma-ray propagation in the presence of Lorentz invariance violation}
The effect of Lorentz invariance violation is expected to be of relevance
at energies  approaching the Planck scale. However, for the propagation
of gamma-rays at TeV energies, the threshold for pair production is shifted leading to an energy dependent and characteristic deviation of the
optical depth at energies above $\approx 10$ TeV.\cite{liv_abs} For some nearby objects like e.g., 
Mkn 501 ($z=0.034$) or Mkn~421 ($z=0.031$), the observed gamma-ray spectra
are sensitive to this effect, observations with future instruments will have the potential to search 
for the spectral imprint with sources at larger distances.\cite{tavecchio} The analysis of gamma-ray spectra presented here
indicates indeed a deviation from the expected absorption. However, the effect is not 
tied to a particular energy scale, but rather to a level of absorption 
($\tau \gtrsim 1$). This is at first sight inconsistent with the 
expectation of a Lorentz-invariance violating effect. However, there are some proposals
in which a more complicated structure of the vacuum would lead to a variation of the
photon dispersion relation depending on the line of sight or distance of the object.\cite{Mavromatos} In this case, such a model may offer an explanation for the observations.
\subsection{Ultra-high energy cosmic-rays and cascading}
 Blazars and flat-spectrum radio quasars are attractive candidates for the sites of 
acceleration of ultra-high energy cosmic-rays (UHECRs), even though neither the
directions of UHECRs nor of energetic neutrinos have so far been found to correlate
significantly with nearby blazars or active galactic nuclei in general.\cite{paoneg} In a scenario, where
blazars are powerful accelerators of UHECRs with a luminosity (in particle acceleration) of $\approx 10^{44}~$ergs/s, the observed
gamma-ray emission may be of secondary origin. While the approach works well in individual cases,\cite{esseykus}
it requires fine-tuning to explain all the data: The transition
of primary to secondary gamma-ray dominated part of the energy spectrum depends only on the ratio of primary gamma-ray luminosity and 
UHECR luminosity. Unless this ratio is tied to the distance in the right way, it would be
difficult to re-produce the observed hardening of the spectrum. 
\section{Conclusions}
The gamma-ray spectra of extra-galactic sources in the energy range from 100 GeV to tens of TeV continue to be of interest in the context of 
propagation in the extra-galactic medium. The standard picture of pair-production absorption in the extra-galactic background light seems to be incomplete.
It does not explain the observations, especially in the tail of the energy spectra observed at optical depth $\tau\gtrsim 1$, where in the case 
of maximum mixing of photons with light axion-like particles, a recovery of the absorbed photon beam is expected to dominate. 
Alternative explanations (including
effects of Lorentz-invariance violation as well as UHECR induced cascades) are attractive possibilities but require additional assumptions/fine-tuning.
Clearly, the future gamma-ray instrumentations, both at the high energy end (e.g., TAIGA-HiSCORE\cite{martint}, LHAASO\cite{lhaaso}) 
as well as at the low energy end (CTA\cite{cta}) will 
provide important observations to clarify the observational situation.

\section*{Acknowledgements}
\footnotesize
The author is thankful for the organizers and the chairs of the session HE1 \textit{Experimental
tests of fundamental physics with high energy gamma-rays} (Alessandro de Angelis and
Razmik Mirzoyan) for adding this topic to the 
very interesting session and to Marco Roncadelli for providing very useful comments on the manuscript.
Without the last-minute help of R.P. Feller, it would not have been
possible to finish this article in time, thank you for the support!

\end{document}